\documentclass[journal, a4paper]{IEEEtran}
\usepackage[T1]{fontenc}
\usepackage[latin9]{inputenc}
\usepackage{color}
\usepackage{amsmath}
\usepackage{amsthm}
\usepackage{amssymb}
\usepackage{esint}
\usepackage{verbatim}
\usepackage{graphicx}
\usepackage{stfloats}
\usepackage{subfigure}
\usepackage{dsfont}
\usepackage{bm}
\usepackage{balance}
\usepackage{multicol}
\usepackage{multirow}
\usepackage{tabularray}
\usepackage{makecell}
\usepackage{algorithm}
\usepackage{algorithmic}
\usepackage[numbers,sort&compress]{natbib}

\makeatletter

\theoremstyle{plain}

\theoremstyle{definition}

\theoremstyle{remark}

\theoremstyle{lemma}

\theoremstyle{assumption}
\newtheorem{assp}{\protect\assumptionname}

\theoremstyle{plain}
\newtheorem{col}{\protect\colloaryname}
\makeatother

\usepackage{babel}
\providecommand{\definitionname}{Definition}
\providecommand{\theoremname}{Theorem}
\providecommand{\remarkname}{Remark}
\providecommand{\lemmaname}{Lemma}
\providecommand{\assumptionname}{Assumption}
\providecommand{\colloaryname}{Colloary}

\begin{document}
\title{Semantic-Aware Power Allocation for Generative Semantic Communications with Foundation Models}


\author{\IEEEauthorblockN{Chunmei Xu,  Mahdi Boloursaz Mashhadi, Yi Ma, Rahim Tafazolli}\\
\IEEEauthorblockA{5GIC \& 6GIC, Institute for Communication Systems, University of Surrey, Guildford, U.K. \\
Emails: \{chunmei.xu; m.boloursazmashhadi; y.ma; r.tafazolli\}@surrey.ac.uk}
}

\maketitle
\thispagestyle{empty}

\begin{abstract} 
Recent advancements in diffusion models have led to a significant breakthrough in generative modeling. The combination of the generative model and semantic communication (SemCom)  enables high-fidelity semantic information exchange at ultra-low rates. In this paper, a novel generative SemCom framework for image tasks is proposed, utilizing pre-trained foundation models as semantic encoders and decoders for semantic feature extraction and  image regeneration, respectively. The mathematical relationship between transmission reliability and the perceptual quality of regenerated images is modeled and the semantic values of extracted features are defined accordingly. This relationship is derived through numerical simulations on the Kodak dataset. Furthermore, we investigate the semantic-aware power allocation problem, aiming to minimize total power consumption while guaranteeing semantic performance.  To solve this problem, two semantic-aware power allocation methods are proposed by constraint decoupling and bisection search, respectively. Numerical results demonstrate that the proposed semantic-aware methods outperform conventional approach in terms of total power consumption.
\end{abstract}


\section{Introduction}
Semantic communications (SemCom)  aim at precise content reconstruction with equivalent semantics, which is fundamentally different from conventional communications targeting accurate source recovering \cite{gunduz2022beyond}.  It has the potential to achieve ultra-low compression rates and extremely high transmission efficiency, which is gaining substantial interest from both academic and industry communities \cite{liang2023generative}. Although efforts to develop semantic information theory have been ongoing since the establishment of Shannon's theory, a comprehensive and universal theory remains elusive.  Nevertheless, the remarkable advancements in artificial intelligence (AI) have paved the way for the development of SemCom systems, particularly in the realm of deep learning-based SemCom. 

The end-to-end architecture is widely used to jointly train the neural network (NN)-based semantic encoder and decoder, facilitating the formation and sharing of the knowledge base between them. The concept of deep joint source and channel coding (JSCC) was first proposed for image tasks by adopting the auto-encoder NN network \cite{bourtsoulatze2019deep}, and its numerous variants were developed subsequently to account for various types of sources and channel models \cite{xie2021deep, weng2021semantic}. These deep JSCC approaches have demonstrated superior performance over the conventional separated source and channel coding schemes in terms of distortion metrics such as mean square error (MSE), peak-signal-to-noise (PSNR), and multi-scale structural similarity (MS-SSIM). However, the distortion may no longer serve as the primary performance indicator for emerging applications with inference goals, where precisely conveying the semantic information is sufficient and more important. To preserve semantic information, the authors in  \cite{erdemir2023generative} proposed to integrate the generative adversarial network (GAN) into a SemCom systems for signal regeneration. This approach significantly outperformed the deep JSCC technique in terms of both distortion and perceptual quality. Recent advancements in state-of-the-art diffusion models have marked a significant breakthrough in generative modeling, demonstrating impressive results in regenerating images \cite{rombach2022high}, audios \cite{ghosal2023text}, and videos \cite{bar2024lumiere}. The diffusion model has been adopted in  SemCom systems for synthesizing semantic-consistent signals, utilizing a loss function combining both MSE and Kullback-Leibler (KL) divergence \cite{grassucci2023generative}. This approach demonstrated high robustness to poor channel conditions and outperformed existing methods in generating high-quality images while preserving semantic information. 

However, adopting end-to-end architectures to learn  deep learning-based SemCom systems present two key practical limitations. First, the employment of analog modulations for data training due to their feasibility and convenience in gradient computation and back-propagation, and the architecture of joint source and channel coding, conflict with modern digital communication systems with open systems interconnection (OSI) model. Secondly, intensive computations are required in the training phase to account for diverse wireless channel characteristics. This focus on specific channel conditions during training potentially results in poor generalization performance.  Concurrently, the field of AI is undergoing a paradigm shift with the emergence of foundation models such as bidirectional encoder representations from transformers (BERT) and generative pre-trained transformers (GPT). These foundation models, trained on vast and diverse datasets, demonstrate the ability to capture general patterns, and thereby form a comprehensive knowledge base. Notably, generative diffusion foundation models such as DALL$\cdot$E  show promise in synthesizing high perceptual quality images with prompt exchanges at ultra-low rates \cite{qiao2024latency}. 

Inspired by these, we propose the generative SemCom framework for image tasks by utilizing powerful pre-trained foundation models to extract semantic features and regenerate signals at the encoder and decoder, respectively. Within this framework, transmission reliability becomes the sole factor influencing the perceptual quality of the regenerated images, with their mathematical relationship modeled as a non-decreasing percpetion-error function.  The semantic values of semantic data streams are defined to measure the semantic information accordingly. We investigate the semantic-aware resource allocation problem in the channel-uncoded case, aiming to minimize the total power consumption while ensuring the perceptual quality of regenerated images. The rest of this paper is organized as follows. Section II introduces the proposed generated SemCom framework for image tasks and defines semantic values. Section III provides the semantic-aware power allocation problem formulation, and Section IV presents the proposed methods. Numerical results are given in Section V to demonstrate the performance of the proposed framework. Finally, Section V concludes this paper.

\section{Generative SemCom Framework}\label{sec:system_model}

\begin{figure*}[tbp]
\centering
\includegraphics[width=0.98\textwidth]{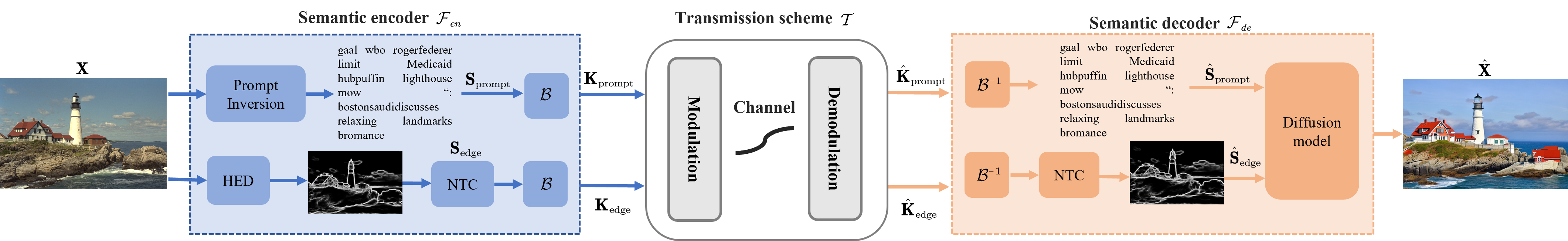}
\caption{The proposed generative semantic communication framework with pre-trained foundation models for image task.}
\label{fig:system_model}
\end{figure*}

The proposed generative  SemCom framework for image tasks, as depicted in Fig. \ref{fig:system_model}, consists of semantic encoder $\mathcal F_{en}$, transmission scheme $\mathcal T$, and semantic decoder $\mathcal F_{de}$.  Before giving a detailed description of the generative SemCom framework, we introduce the semantic metric based on contrastive language-image pre-training (CLIP) similarity  \cite{radford2021learning} to evaluate the perceptual quality of the regenerated image. The metric is written as
\begin{equation}
\label{eq:clip}
P\triangleq \mathbb E \left[ \mathrm{CLIP}\left(\mathbf{X},\hat{\mathbf{X}}\right)\right]= 1-\mathbb E \left[\frac{F_\mathrm{clip}\left(\mathbf{X}\right)\cdot F_\mathrm{clip}\left(\hat{\mathbf{X}}\right)}{\big\Vert F_\mathrm{clip}\left(\mathbf{X}\right)\big\Vert \big\Vert F_\mathrm{clip}\left(\hat{\mathbf{X}}\right)\big\Vert }\right],
\end{equation}where  ${\mathbf X}$ and $\hat{\mathbf X}$ denote the source and the regenerated images, respectively. $F_\mathrm{clip}\left(\cdot\right)$ refers to a pre-trained model trained on a large text-image dataset \cite{radford2021learning}.

\subsection{Semantic Encoder}
The source image is encoded into two distinct semantic features, namely the textual prompt and the edge map features, utilizing two semantic extractors based on pre-trained foundation models.  The textual prompt is extracted by textual transform coding via prompt inversion \cite{wen2023hard}. The edge map feature is extracted using the Holistically-nested Edge Detection (HED) with a non-linear transform code (NTC) model  \cite{balle2020nonlinear}  for further compression. For notional simplicity, we use subscripts $1$ and $2$ to replace subscripts $\mathrm{prompt}$ and $\mathrm{edge}$ in the sequel. 
The  $i$th extracted feature can be expressed by 
\begin{equation}
\mathbf{S}_{i}=F_{en,i}\left(\mathbf{X}\mid \bm{\theta}_i^*\right), \forall i \in\{1,2\}
\end{equation}where $F_{en,i}\left(\mathbf{X}\mid \bm{\theta}_i^*\right)$ is the $i$th pre-trained foundation model with $\bm{\theta}_i^*$ being the NN parameters. 

To ensure compatibility with  existing digital communication systems, the semantic feature $\mathbf{S}_i$ is converted into the bit sequence denoted as $\mathbf{K}_{i}$. We have 
$\mathbf{K}_{i}=\mathrm{\mathcal{B}}\left(\mathbf{S}_{i}\right)$, where $\mathrm{\mathcal{B}}\left(\cdot\right)$ is a binary mapping function such as ASCII, Unicode encoding and quantization.  In SemCom systems, the semantic data streams contribute unequally to the perceptual quality of the regenerated image. This varying contributions can be measured using a semantic metric closely related to the inference goal or task at the receiver. 
This is fundamentally different from conventional communication systems. Denote the semantic value of the $i$th semantic data stream as $L_i$ to quantify its semantic information in terms of a specific semantic metric. Generally, the semantic data stream with a larger $L_{i}$ has a greater impact on the perpetual quality of the regenerated signal, indicating its greater importance.  

\subsection{Transmission Scheme} 

Due to the different importance of the semantic data streams, multi-stream transmissions are considered in the proposed generative SemCom framework.  The received data streams are expressed as 
\begin{equation}
\left[\hat{\mathbf{K}}_{1},\hat{\mathbf{K}}_{2}\right]=\mathcal{T}\left(\left[\mathbf{K}_{1},\mathbf{K}_{2}\right]\right),
\end{equation}
where $\mathcal{T}\left(\cdot\right)$ is the transmission scheme, which may comprise the channel coding/decoding and modulation/demodulation components. The semantic data streams are considered to be transmitted in an orthogonal manner to mitigate the inter-stream inference. Despite this, errors may still occur in the received semantic data stream $\hat{\mathbf{K}}_i$  due to the fading and noisy effects of the wireless channels. 

\subsection{Semantic Decoder}
In the semantic decoder,  the pre-trained generative foundation model $F_{de}$, i.e., ControlNet  \cite{zhang2023adding} built upon the Stable Diffusion model \cite{rombach2022high},  is employed to synthesize the received semantic data streams into an image  $\hat{\mathbf{X}}$. In the channel-uncoded case, the received data streams regardless of transmission errors are processed by the generative foundation for signal synthesizing, as the transmission errors cannot be identified and corrected. The received semantic data streams $\hat{\mathbf{K}}_{i}$ are first reconverted into the semantic features $\hat{\mathbf{S}}_i=\mathcal{B}^{-1}\left(\hat{\mathbf{K}}_{i}\right)$ with $\mathrm{\mathcal{B}^{-1}}\left(\cdot\right)$ being the inverse operation of $\mathrm{\mathcal{B}}\left(\cdot\right)$.  $\hat{\mathbf{S}}_i$ are forwarded to the generative foundation model $F_{de}$ to  synthesize  $\hat{\mathbf{X}}$, which can be expressed as
\begin{equation}
\hat{\mathbf{X}}=F_{de}\left(\hat{\mathbf{S}}_{1},\hat{\mathbf{S}}_{2}\mid \bm{\omega}^*\right)=\mathcal F_{de}\left(\hat{\mathbf{K}}_{1},\hat{\mathbf{K}}_{2}\right),
\end{equation}where $\bm{\omega}^*$ are the NN parameters of the generative diffusion model.  Denote $\hat{L}_{i}$  as the semantic values of the $i$th received semantic data stream $\hat{\mathbf{K}}_i$.  The semantic information is lossy due to the transmission errors, thus we have $\hat{L}_{i}\leq L_i$. 

Given the semantic encoder and decoder, the transmission scheme and wireless channels remain to influence the perceptual quality of the regenerated image. As a consequence, the transmission reliability becomes the sole factor impacting the achieved perceptual quality. Denoting the bit error rate (BER) of the $j$th bit of $\hat{\mathbf{K}}_i$ as $\psi_{ij}$, the perception value $P$ defined in (\ref{eq:clip}) becomes a function of $\psi_{ij}$.
\begin{assp}\label{assp1}
Assume that the perception value  $P$ is non-decreasing with respect to (w.r.t.) the BER $\psi_{ij}$.
\end{assp}

The semantic value of $i$th transmitted semantic data stream $\mathbf{K}_i$ is defined as 
\begin{equation}
    L_i = 1-\underline{P}_i,
\end{equation}where $\underline{P}_i$ is the perception value of regenerated signal $\hat{\mathbf{X}}_{i}^*=\mathcal F_{de}\left(\mathbf{K}_i\right)$ synthesized only by the $i$th semantic data stream $\mathbf{K}_i$. For the received semantic data stream $\hat{\mathbf{K}}_i$, the semantic value is defined as 
\begin{equation}
\hat{L}_{i}\left(\left\{\psi_{ij}\right\}_{j}\right)= 1-P_{i}\left(\left\{\psi_{ij}\right\}_{j}\right), 
\end{equation}
where $P_{i}\left(\left\{\psi_{ij}\right\}_{j}\right)$ is the perception value of $\hat{\mathbf{X}}_{i}=\mathcal F_{de}(\hat{\mathbf{K}}_i)$ synthesized only by $\hat{\mathbf{K}}_i$.  

\section{Problem Formulation of Semantic-Aware Power Allocation }
Transmission reliability significantly affects both the perceptual quality of the regenerated images and the resource consumption. Unlike conventional communication systems that treat all data streams equally, SemCom systems offer the opportunity to exploit the semantic importance to enhance resource efficiency.  In this paper, we investigate the semantic-aware power allocation problem for the proposed generative SemCom framework. The objective is to minimize total power consumption while guaranteeing semantic performance.    

 Let $z_{i}$ be the transmitted symbol of the $i$th  semantic data stream with unit energy  such that $\mathbb{E}\left\{ z_{i}z_{i}^{\mathrm{H}}\right\} =1$. The $i$th received semantic signal can be written as 
\begin{equation}
y_{i}=\sqrt{q_{i}}h_iz_{i}+n_{i},
\end{equation} where $h_{i}$ is channel assumed to be quasi-static and modelled as $h_i = \sqrt{h_0\left(\frac{d}{d_0}\right)^{-\alpha}}\Tilde{h}_i$ where $h_0\left(\frac{d}{d_0}\right)^{-\alpha}$ is the path loss at distance $d$ with $h_0$ being the path loss at reference distance $d_0$. $\Tilde{h}_i$ is the Rayleigh fading channel with a covariance of $1$ and   $n_{i}$  is the Gaussian noise following the distributions of $n_{i}\sim\mathcal{CN}\left(0,\sigma_{i}^{2}\right)$. $q_{i}$ is the allocated power for each symbol of the $i$th semantic data stream. 

Under the quasi-static channel, the received signal-to-noise ratio (SNR) of  each symbol is equal, which is given by
\begin{equation}
\mathrm{SNR}_{i}=\frac{q_{i}\vert h_i\vert^2}{\sigma_i^2}.
\end{equation}The BER of each bit of the $i$th semantic data  is given by 
\begin{equation} 
\psi_{i} = \frac{a_i}{\log_2M_i}Q\left(\sqrt{b_i\mathrm{SNR}_i}\right),\label{eq:BER_p}
\end{equation}where $Q\left(x\right)=\frac{1}{\sqrt{2\pi}}\int_x^{\infty}e^{(-\frac{u^2}{2})}du$ is the Q-function.  Parameters $a_i$ and $b_i$ depend on the adopted modulation scheme. $M_i$ is the modulation order.  The BER of different modulation schemes are
listed in \cite[Table 1]{simon2001digital}. The problem, which minimizes the total power consumption while ensuring the semantic performance $\bar{P}$ under the channel-uncoded case, can be formulated as
\begin{subequations}
\begin{align}
(\mathcal P1): \quad \min_{q_{i}}\quad & \sum_{i=1}^I K_iq_{i}  \label{eq:power_obj}\\
\mathrm{s.t.\quad} &  
P\left(\left\{\psi_{i}\right\}_i\right)\le\bar{P}\label{eq:power_allocation_ber_f_cons1}
\end{align}\label{eq:problem1}\end{subequations}
 To solve the problem, the following corollary is established according to \textbf{Assumption} \ref{assp1}, since the BER $\psi_{i}$ is monotonically decreasing with the allocated power $q_{i}$.
\begin{col}\label{lem4}
    The optimal solution $q_i^*$ to problem $\mathcal P1$   satisfies the equality of  constraint \rm{ (\ref{eq:power_allocation_ber_f_cons1})}. 
\end{col}

\section{Semantic-Aware Power Allocation Methods}
This section presents two semantic-aware power allocation methods, namely the semantic-aware proportional method and semantic-aware bisection method. 


\subsection{Semantic-Aware Proportional Method}
Assuming the independence of semantic data streams,  the constraint (\ref{eq:power_allocation_ber_f_cons1}) can be decoupled into $I$ independent constraints, each corresponding to an individual semantic value constraint. Problem ($\mathcal P1$)  is then relaxed  into 
\begin{subequations}
\begin{align}
(\mathcal P2):\quad \min_{q_{i}}\quad & \sum_{i=1}^I K_iq_{i} \\
\mathrm{s.t.\quad} &\hat{L}_{i}\left(\psi_{i}\right) \ge \bar{L}_i,\,\, \forall i\in\mathcal I,\label{eq:P1-1Cons1},
\end{align}
\end{subequations}
where $\bar{L}_i$ is the semantic value requirement of the $i$th received semantic data stream corresponding to the semantic performance requirement $\bar{P}$.  Based on \textbf{Assumption \ref{assp1}},  the semantic value of the received semantic data stream is non-increasing w.r.t. the BER $\psi_{i}$. Therefore, the optimal solutions to $\mathcal P2$ are obtained when the equalities of constraints (\ref{eq:P1-1Cons1}) hold. Denoting $\psi_{i}^*$ as the solution obtained by solving equation $\hat{L}_{i}\left(\psi_{i}\right) = \bar{L}_i$, the optimal solutions can be readily obtained by  substituting $\psi_{i}^*$  back to (\rm{\ref{eq:BER_p}}), which is given by
    \begin{equation}\label{eq:proportional_p}
     \quad q_{i}^* = \frac{\sigma_{i}^{2}}{b_i\vert h_{i}\vert^2}\left(Q^{-1}\left(\frac{\log_2M_i}{a_i}\psi_{i}^*\right)\right)^2,
    \end{equation} 

\subsection{Semantic-Aware Bisection Method}

\begin{algorithm}[t]
	\renewcommand{\algorithmicrequire}{\textbf{Input:}}
	\renewcommand{\algorithmicensure}{\textbf{Output:}}
	\caption{Semantic-aware bisection  method} 
	\label{alg1}
	\begin{algorithmic}[1]

		\STATE Initialization:   $\left(\psi^L_{1},\psi^L_{2}\right)$,  $\left(\psi^R_{1},\psi^R_{2}\right)$
         \STATE \textbf{while} $\psi^R_{1}-\psi^L_{1} \ge \epsilon$
		\STATE \quad $\psi_1=(\psi^R_{1}+\psi^L_{1})/2$
  
		\STATE \quad  Obtain $\psi_2$ by solve the equation (\ref{eq:pro1-2_cons1}) 

            \STATE  \quad Compute partial gradients     $\left(\frac{\partial f}{\partial \psi_1}, \frac{\partial f}{\partial \psi_2}\right)$
            
            \STATE  \quad Compute gradient  $\nabla_{\psi_1}\psi_2$  by implicit differentiation of (\ref{eq:pro1-2_cons1})            
             \STATE  \quad \textbf{if} $\frac{\partial f}{\partial \psi_1}+\nabla_{\psi_1}\psi_2\frac{\partial f}{\partial \psi_2}\ge 0$
             
            \STATE \qquad$\left(\psi^R_{1}, \psi^R_{2}\right)\leftarrow \left(\psi_1, \psi_2\right)$
            
            \STATE  \quad \textbf{else} 
            \STATE  \qquad$\left(\psi^L_{1}, \psi^L_{2}\right)\leftarrow \left(\psi_1, \psi_2\right)$
            \STATE  \quad \textbf{end} 
		\STATE  \textbf{end}
	\end{algorithmic}  
\end{algorithm}
Based on \textbf{Corollary \ref{lem4}}, problem $\mathcal P1$ can be reduced into 
\begin{subequations}
\begin{align}
(\mathcal P3): \quad \quad\min_{\psi_{1}, \psi_{2}}\quad & \sum_{i=1}^{2}\frac{K_i\sigma_{i}^{2}}{b_i\vert h_{i}\vert^2}\left(Q^{-1}\left(\frac{\log_2M_i}{a_i}\psi_{i}\right)\right)^2\\
\mathrm{s.t.\quad} &  P\left(\psi_{1}, \psi_{2}\right)=\bar{P}.\label{eq:pro1-2_cons1}
\end{align}\label{eq:problem0}\end{subequations}
The feasible solutions $\left(\psi_{1}, \psi_{2}\right)$ form a line on the perception-error surface. For any two feasible solutions $\left(\psi^{(1)}_{1}, \psi^{(1)}_{2}\right)$ and $\left(\psi^{(2)}_{1}, \psi^{(2)}_{2}\right)$,  we have $\psi^{(2)}_{2}\le\psi^{(2)}_{2}$ if $\psi^{(1)}_{1}\ge\psi^{(2)}_{1}$.  The main idea is to find the solution with the gradient of the objective function being $0$, which is obtained by the bisection search technique. 
  Denoting the two ends of the line as $\left(\psi^L_{1},\psi^L_{2}\right)$ and $\left(\psi^R_{1},\psi^R_{2}\right)$  where $\psi^R_{1}\ge\psi^L_{1}$, the procedure to obtain the solution is summarized in \textbf{Algorithm \ref{alg1}}.

\section{Numerical Results}
To transmit the textual prompt and edge map data streams, the communication parameters are set as follows. The modulations of these two semantic data streams are the same.  Both 8-QAM and 16-QAM modulation schemes are considered.  The channel parameters are set to  $d=100$ m, $d_0=1$ m, $h_0=-30$ dB, and $\alpha=-3.4$. The noise power is set to $\sigma_i^2=-110$ dBm. 

\begin{figure}[th]
\centering
\subfigure[Kodim01: The BPP is $0.0278$]{
\includegraphics[width=0.45\textwidth]{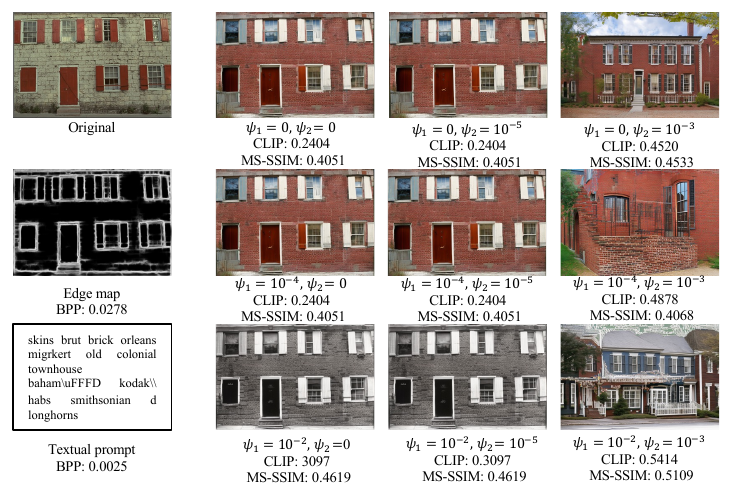}
}
\subfigure[Kodim21: The BPP is $0.02597$]{
\includegraphics[width=0.44\textwidth]{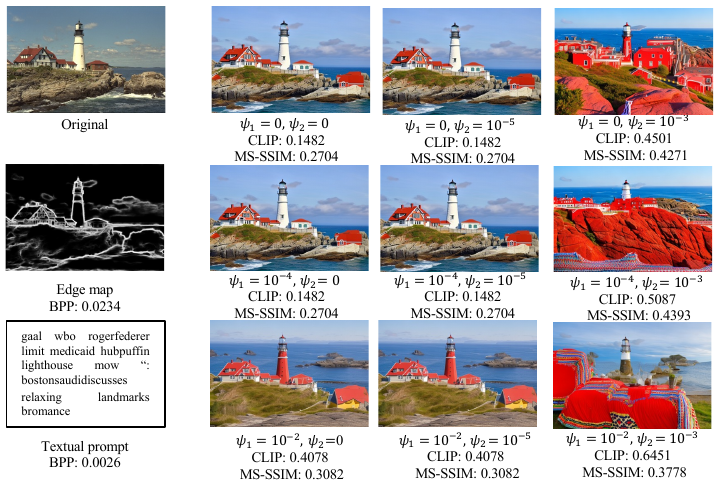}}
\caption{The visual qualities of regenerated images via the proposed generative SemCom System. }
\label{fig:examples}
\end{figure}

\begin{figure}[t]
\centering
\includegraphics[width=0.45\textwidth]{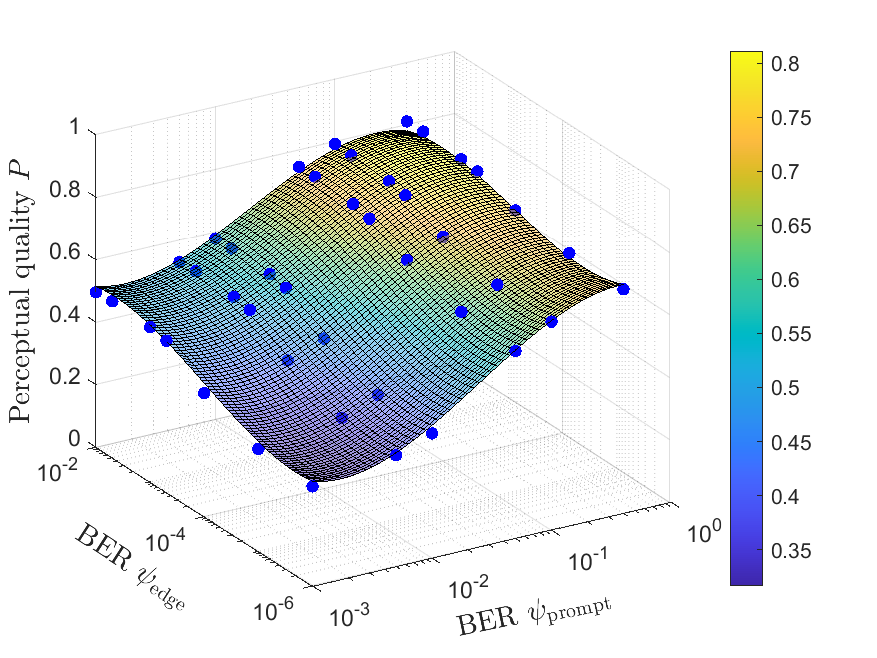}
\caption{The perception-error functions based on Kodak dataset in terms of the CLIP metric. }
\label{fig:P_E_function}
\end{figure}

\begin{figure}[tbp]
\centering
\includegraphics[width=0.4\textwidth]{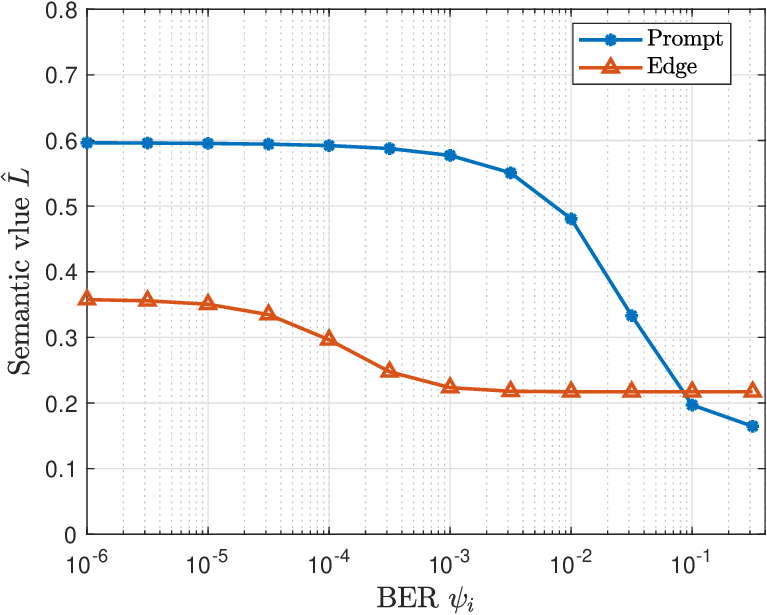}
\caption{Semantic values of textual prompt and edge map semantic data streams based on Kodak dataset in terms of CLIP metric.}
\label{fig:semantic values}
\end{figure}

Fig. \ref{fig:examples}  depicts two image examples using the proposed generative SemCom framework to demonstrate the achieved perceptual quality. As the increase of the BERs, the semantic performance in terms of the CLIP metric is degraded.   The compression rates obtained are $0.0278$ and $0.02597$ bits per pixel (BPP), indicating that ultra-low rates can be achieved within the proposed generative SemCom framework.   It is difficult to explicitly obtain the mathematical relationship between the BERs and the perceptual quality of the regenerated images. Instead, we conduct numerical simulations on the Kodak dataset \cite{Kodak} to empirically derive this function. As shown in Fig. \ref{fig:P_E_function}, the perception-error function is non-decreasing with BERs $\psi_i$, which is obtained by curve fitting using the numerical simulation points. Fig. \ref{fig:semantic values} depicts the defined semantic values of both transmitted and received semantic data streams. The semantic values of textual prompt and edge map streams are $L_1=0.5887$ and $L_2=0.3596$, respectively. For the received semantic data streams, their semantic values, i.e., $\hat{L}_1$ and $\hat{L}_2$,  are non-increasing with BERs $\psi_i$. In addition, the prompt feature has a greater impact on the CLIP performance compared to the edge map feature. However, the edge map feature exhibits greater vulnerability to the BER than the prompt feature due to its larger data length. 

\begin{figure}[tbp]
\centering
\includegraphics[width=0.4\textwidth]{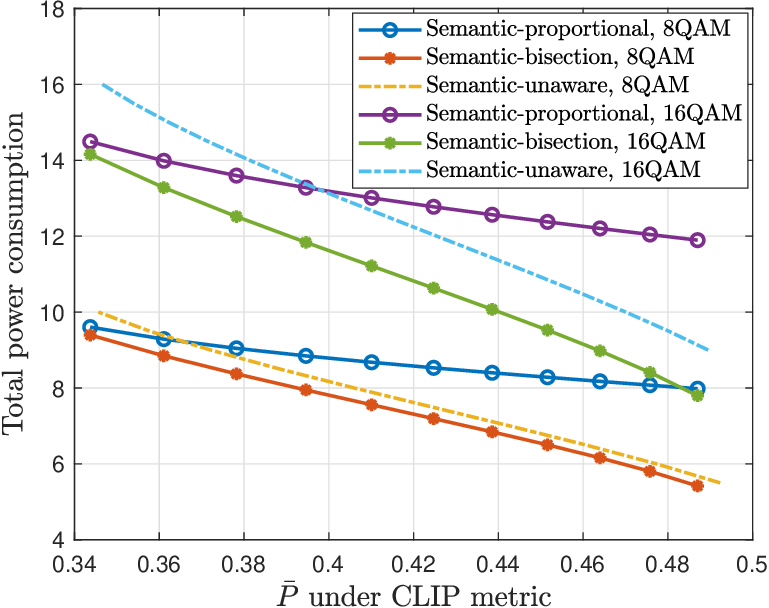}
\caption{Total power consumption versus the perceptual performance requirement $\bar P$ in terms of the CLIP metric.}\label{fig:power_barP}
\end{figure} 

The proposed semantic-aware proportional and bisection methods are compared with the conventional semantic-unaware one that  treats all data streams equally. For the semantic-unaware method, the SNRs for both semantic data streams are the same. For the semantic-proportional method, the allocated power is obtained based on (\ref{eq:proportional_p}), where $\frac{\hat{L}_1}{L_1}=\frac{\hat{L}_2}{L_2}$. The total power consumption comparison results are given in Fig. \ref{fig:power_barP}, showing that the total power consumption decreases as the increase of the performance requirement $\bar{P}$. Under stringent semantic performance requirements, the semantic-proportional method consumes lower power than the conventional approach. However, this performance advantage diminishes as  $\bar{P}$ increases.  The proposed semantic-aware bisection method consistently outperforms the semantic-aware proportional and the semantic-unaware methods. Moreover,  it can be observed that  higher modulation orders lead to increased power consumption due to lower transmission reliability. Notably, the performance advantage of the proposed semantic-aware methods over the semantic-unaware one becomes more evident as the increase of modulation order. 

\section{Conclusion}
A generative SemCom framework for image tasks was proposed in this work, leveraging pre-trained foundation models for both semantic encoding and decoding.  Given the semantic encoder and decoder,  the transmission reliability emerged as the primary factor influencing the perceptual quality of the regenerated images. Their mathematical relationship was modeled as a perception-error function, and the semantic values of the semantic data streams were defined accordingly. Both the perception-error function and semantic values were empirically derived through numerical simulations on the Kodak dataset, providing a quantitative basis for further optimization.  We investigated the semantic-aware power allocation problem and proposed semantic-aware proportional and semantic-aware bisection methods. Numerical results demonstrated that the proposed semantic-aware bisection method consistently outperformed the semantic-aware proportional method and the conventional approach. The performance advantages of the proposed semantic method become more pronounced with the increase of modulation order.

\section{Acknowledgement}
This work was supported by the U.K. Department for Science, Innovation, and Technology under Project TUDOR (Towards Ubiquitous 3D Open Resilient Network).

\bibliographystyle{IEEEtran}
\balance
\bibliography{semantic}

\end{document}